\documentclass{jfm}
\usepackage{graphicx}
\usepackage{epstopdf, epsfig}
\usepackage{newtxtext}
\usepackage{newtxmath}
\usepackage{natbib}
\usepackage{siunitx}
\usepackage{hyperref}
\usepackage{subfigure}
\usepackage{float}

\shorttitle{Guidelines for authors}
\shortauthor{A. N. Other, H.-C. Smith and J. Q. Public}

\title{Learn to Flap: Foil Non-parametric Path Planning via Deep Reinforcement Learning}

\author{Z.P. Wang\aff{1,2,*},
 R.J. Lin\aff{3,4,*}\footnotetext{* These authors contribute equally.}, 
 Z.Y. Zhao\aff{3,4},
P.M. Guo\aff{1,+},
N. Yang\aff{3,4,+}\and
D.X. Fan\aff{1}\corresp{\email{fandixia@westlake.edu.cn}}
}

\affiliation{\aff{1} School of Engineering, Westlake University, Hangzhou 310024, Zhejiang, China.
\aff{2}School of Mechanical and Material Engineering, Queen’s University, Kingston, Ontario K7L 3N6, Canada. 
\aff{3}University of Chinese Academy of Sciences, Beijing 100049, China
\aff{4}Institute of Automation, Chinese Academy of Sciences, Beijing 100190, China}

\begin{document}

\maketitle

\begin{abstract}
To optimize flapping foil performance, the application of deep reinforcement learning (DRL) on controlling foil non-parametric motion is conducted in the present study. Traditional control techniques and simplified motions cannot fully model nonlinear, unsteady and high-dimensional foil-vortex interactions.
A DRL-training framework based on Proximal Policy Optimization and Transformer architecture is proposed. The policy is initialized from the sinusoidal expert display. We first demonstrate the effectiveness of the proposed DRL-training framework which can optimize foil motion while enhancing foil generated thrust. By adjusting reward setting and action threshold, the DRL-optimized foil trajectories can gain further enhancement compared to sinusoidal motion.
Via flow analysis of wake morphology and instantaneous pressure distributions, it is found that the DRL-optimized foil can adaptively adjust the phases between motion and shedding vortices to improve hydrodynamic performance. 
Our results give a hint for solving complex fluid manipulation problems through DRL method.
\end{abstract}


\section{Introduction}
\label{sec:intro}

Designing novel bio-inspired flapping propulsors has been of great interest to the scientific community, as ``about $10^9$ years of animal evolution ... have inevitably produced rather refined means of generating fast movement at low energy cost" \citep{lighthill1969hydromechanics}. For simplification, researchers mainly use the model of sinusoidally flapping foils whose trajectories can be easily parameterized \citep{wu2020review}. For example, studies show that the flapping efficiency is strongly associated with the Strouhal number $Sr$, manifested in its narrow range for the optimal propulsive strategy of aquatic animals across different sizes \citep{qi2022recent}. However, recent observations reveal animals travel in non-sinusoidal motions, such as the Burst-and-Coast pattern of fish \citep{li2021burst}, to manipulate the flow and achieve higher efficiency \citep{chin2016flapping} and better maneuverability \citep{triantafyllou2016biomimetic}.

As the flapping motion becomes more complex, its trajectory cannot be easily parameterized. Preliminary studies \citep{teng2016effects,LiuPRE2020,2021Burst} sampled several non-sinusoidal flapping trajectories and showed a hydrodynamic enhancement that cannot be ignored. In fact, even a tiny change in the instantaneous angle of attack may result in significant force alternation, attributed to the complex foil-vortex interactions \citep{izraelevitz2014adding}. Therefore, the underlying flow mechanism of foil non-parametric flapping has yet to be fully explored and remains unclear. As a typical nonlinear, unsteady, and high-dimensional flow control problem, optimizing such non-parametric flapping trajectories poses a considerable challenge for traditional control techniques that may result in inefficient and intractable solutions.

Deep Reinforcement Learning (DRL) recently has gained significant attention in fluid mechanics for its astonishing achievement in solving complex problems of games \citep{silver2017mastering}, robotics \citep{won2020adaptive} and other industrial tasks \citep{degrave2022magnetic}. Several successes include reducing bluff body drag \citep{rabault2019artificial,fan2020reinforcement} and enhancing airfoil lift \citep{wang2022deep}, by learning statistically mean control actions to induce a switch to a favourable wake pattern. However, as mentioned above, optimizing foil flapping motions requires learning a coherently cyclical motion, meticulously manipulating the strength and timing of the shedding vortices, and their interaction with the moving foil \citep{muhammad2022efficient}, which a simple transfer of the readily available DRL algorithm may not work.

The present study proposes a DRL training framework based on Proximal Policy Optimization (PPO) algorithm and Transformer architecture. In addition, the policy is initialized from the expert display instead of randomization. We compare with other DRL frameworks to demonstrate whether the proposed agent can learn to flap. By carefully adjusting the reward optimization weights and motion thresholds, we illustrate considerable improvements in flapping hydrodynamic performance with learned non-parametric flapping trajectories. At last, by flow visualization, we shed light on why the agent may flap better compared to the statistically equivalent sinusoidal motion.
 
\section{Materials and Methods}
\label{sec:2}
\subsection{Physical model}
We numerically study a 2-dimensional NACA0016 foil flapping in the uniform inflow at $Re = \frac{U_\infty c}{\nu}$ = 1173, where $U_\infty$ is the uniform free stream velocity, $c$ is the foil chord length and $\nu$ is the fluid kinematic viscosity. The trajectory of the flapping foil is first prescribed as a sinusoidal motion combined with both heave $h_s(t)$ and pitch $\theta_s(t)$ around $c/4$, and is later learned intelligently by DRL agents. The prescribed sinusoidal motion can be parameterized as follows,
\begin{equation}\label{eqn1}
    h_s(t)=h_0\sin(2\pi ft), \quad\theta_s(t) =\theta_0\sin(2\pi ft+\phi),
\end{equation}
where the frequency $f$ is set to be the same for both pitch and heave with amplitudes of $\theta_0$ and $h_0$, respectively. And $\phi$ denotes the phase difference between the two motions. Therefore, the non-dimensional parameters of Strouhal number $Sr$ and scaled amplitude factor $A_D$ can be defined as follows, 
\begin{equation}
     \quad Sr = \frac{fD}{U_\infty},\quad A_D = \frac{2A}{D},
 \end{equation}
where $D$ is the foil thickness, and $A$ is the foil peak-to-peak trailing edge (TE) amplitude. And we quantify the flapping performance via the mean thrust coefficient $\bar{C_T}$ and efficiency coefficient $\eta$ as follows, 
\begin{equation}
    \bar{C}_T = \frac{1}{\mathcal{T}}\int_0^{\mathcal{T}}C_T dt, \quad
    \eta=\frac{\bar{C}_T}{\int_0^{\mathcal{T}}C_{\mathcal{P}} dt}=\frac{\bar{C}_T}{\int_0^{\mathcal{T}}\frac{1}{U_\infty}(C_L\dot{h}+C_M\dot{\theta}) dt},
\end{equation}

\noindent where $C_T = \frac{2F_x}{\rho U_\infty ^2 c}$, $C_L = \frac{2F_y}{\rho U_\infty ^2 c}$ 
and $C_M = \frac{2M}{\rho U_\infty ^2 c^2}$ are the instantaneous thrust, lift and moment coefficients. $F_x$ and $F_y$ are the fluid forces opposite and perpendicular to the inflow direction. $M$ is the fluid moment around the pitching point. $\dot{h}$ and $\dot{\theta}$ are the heaving and angular velocity respectively. $\rho$ is fluid density, and $\mathcal{T}$ is the flapping period.  

For the DRL learning cases, the instantaneous heaving and pitching velocities $\hat{V}_y$ and $\hat{V}_\theta$ are used, and hence the instantaneous heaving position and pitching angle are cumulative values from the start of training till the current time, which can be described by:
\begin{equation}\label{eqn2}
    y_{l} (t)= \sum_{i=0}^t \hat{V}_{y}^i \cdot \delta t,\enspace
    \theta_{l} (t) = \sum_{i=0}^t \hat{V}_{\theta}^i \cdot \delta t,
\end{equation}
where $i$ denotes the current step for simulation. The multiplier $\delta t$ is one time step in simulation.

\begin{figure}
\centerline{\includegraphics[width=0.9\textwidth]{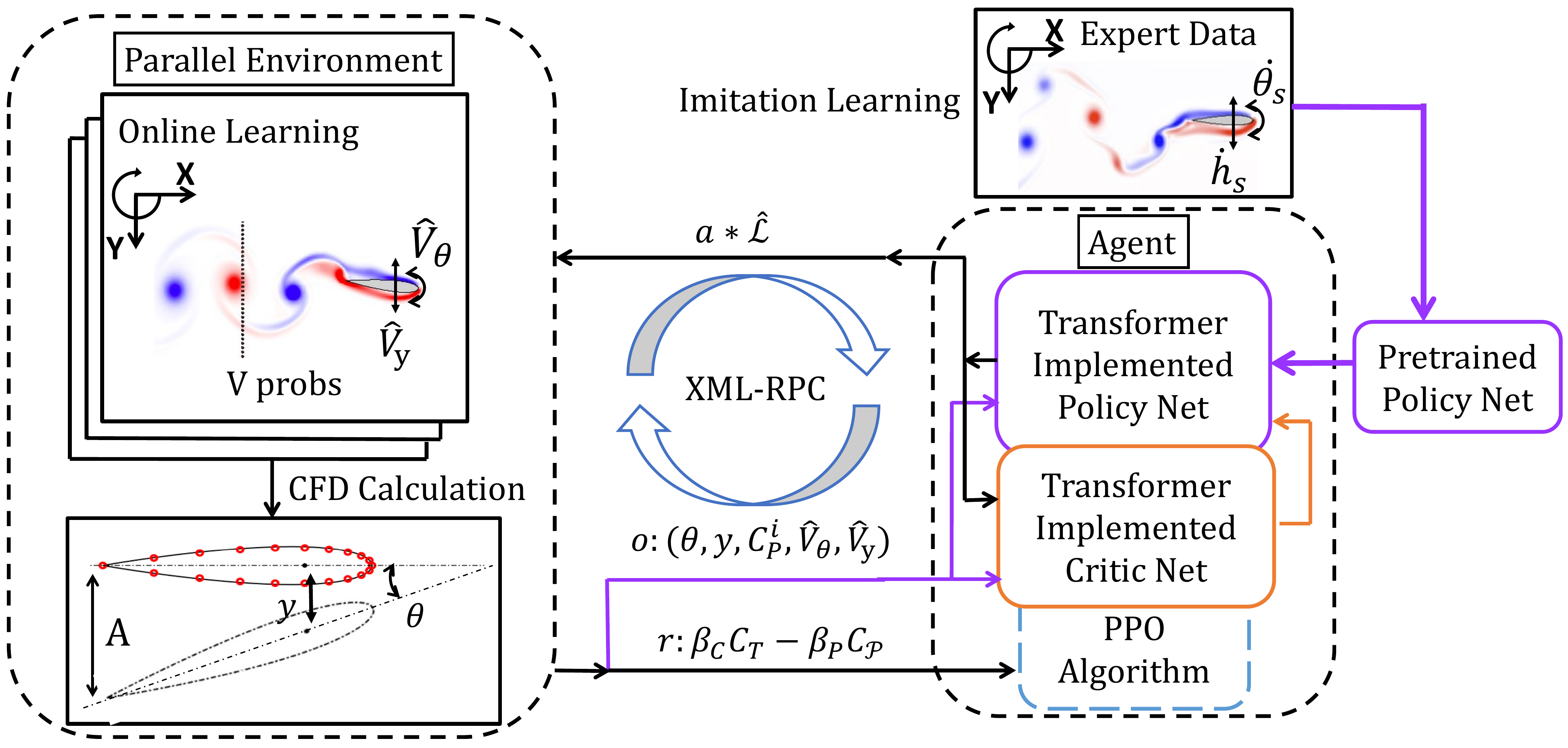}}
\caption{Sketch of RL framework and data flow. The arrows indicate the control sequence: the policy net is first pre-trained by the expert data (selected sinusoidal motion). The agent inquires about the states via observation $o$ from ten parallel environments, then sends actions $a$ adjusted by $\hat{\mathcal{L}}$. We implement XML-RPC protocol to enable cross-platform communication between the environment (CFD solver) and RL agent (python).}
\label{fig:DRL framework}
\end{figure}

\subsection{Numerical Method}

The CFD solver based on the boundary data immersion method (BDIM)\citep{weymouth2011boundary} was chosen in the current work for its capability to simulate complex geometries undergoing rapid motion with large amplitudes \citep{schlanderer2017boundary}. The solver has been extensively validated with various experimental data \citep{maertens2015accurate}. In detail, the simulation is set the resolution of 32 grids per foil chord and a domain size of $16c \times 12c$. The calculation time step is set to adapt dynamically to the complexity of the calculation. The calculation speed and efficient data communication make current BDIM solver a potential DRL environment. 

It is noted that both the mesh density, domain size, Reynolds number and resolution could easily be increased in the future study, but are kept low here as it allows for fast training which is the primary aim to demonstrate our proof-of-concept of the DRL learning process for flapping foil.

\subsection{Reinforcement Learning Framework and Algorithm}\label{sec:2.3}

\textbf{Key Challenges.}  Integrating DRL into foil non-parametric path planning entails overcoming domain-specific obstacles, such as partial observation limitations, vast exploration spaces, and multi-objective preferences. To address these challenges,  we employ a Proximal Policy Optimization (PPO)-based algorithm \citep{ppo} combined with a transformer architecture \citep{transformer}. Since limited environmental data (e.g., force, pressure on the foil) can be obtained in real-world situations, we formulate this issue as a Partially Observable Markov Decision Process (POMDP), effectively addressed through our Transformer architecture implementation.  In the following, we emphasize the primary elements of our DRL framework. More detailed descriptions of PPO, POMDP and Transformer will be given in Appendix A. 

\textbf{PPO algorithm. } PPO has several advantages, such as facilitating support for large-scale parallel training, essential for long-episodic and high-dimensional tasks \citep{openfive}. In addition, PPO is robust as an efficient approximation of the trust region optimization approach, which promises reliable policy improvement in noisy environments. Furthermore, the use of Generalized Advantage Estimation (GAE) assists in long-term credit assignment, improving the algorithm's performance and data efficiency.

\textbf{Transformer. } In order to effectively model the time-dependent behaviour of foil flapping, we employ the Transformer architecture, which has been demonstrated to excel in capturing long-term interactions and supporting high training throughput  \citep{brown2020gpt3, transformerQ}. Central to the Transformer is its attention mechanism, which uncovers intricate inter-dependencies within input sequences through a scaled dot-product function. 

\textbf{Pretraining on Expert Demonstration. } The foil action's cause-and-effect relationship is non-instantaneous and the complete motion pattern consists of thousands of steps. This vast policy space enables the potential discovery of superior foil motion control strategies. However, the exponentially-expanding exploration space as a function of the simulation length poses challenges to the learning performance. To address the substantial exploration space, we use a pre-trained model, imitated from sinusoidal expert policies, as an initial starting point in the high-value subspace of the overall policy space. 

\textbf{Diverse Reward Functions.} Foil motion optimization objectives are efficiency coefficient and thrust coefficient thus designing diverse reward functions to boost the diversity of motion patterns is essential. The reward function applied in the present work is 
\begin{equation}
    r_t=\beta_C \text{clip}(C_T^t, -C_{T}^{m}, C_{T}^{m}) - \beta_\mathcal{P} \text{clip}(C_{\mathcal{P}}^t, - C_{\mathcal{P}}^{m}, C_{\mathcal{P}}^{m})
    \label{eq:r}
\end{equation}
to balance the maximization of thrust and energy consumption of foil motion. The clip parameters $C_{T}^{m}$, $C_{\mathcal{P}}^{m}$ alleviate extreme value of $C_T$ and $C_{\mathcal{P}}$, and linear weight $\beta_C, \beta_P$ balance the importance between thrust and energy consumption. 
The tuple $(\beta_C$, $\beta_\mathcal{P}$, $C_T^m$, $C_{\mathcal{P}}^m)$ describes a specific reward function, which is initialized as $(0.1, \frac{1}{3000}, 10, 3000)$ respectively.

\textbf{Training Pipeline.} To enhance data interaction throughput, we employ ten parallel simulations, serving as the environment to interact with the transformer agent to minimize training time. Communication sequences and data flow are shown in figure \ref{fig:DRL framework}. Agents are initialized from a fully developed simulation in which the foil remains stationary and void of DRL interference, ensuring a stable vortex shedding state at the onset of training. The pretrained Transformer receives 24 observation signals, normalized to $[0,1]$, and outputs normalized rotational and vertical velocity, $\hat{V}_\theta$ and $\hat{V}_y$. These output actions are then scaled by $\hat{\mathcal{L}}$ for amplitude adjustment. Initialized at $[0.5,0.5]$, $\hat{\mathcal{L}}$ remains constant throughout training, and its effects will be discussed in Section 3.2. The interactions and updates will be repeated at every time step between agents and the environment.

\section{Results and discussion}
\label{sec:3}

We divide this section into three parts, each designed to examine a particular aspect of the DRL-based control flapping foil performance: 1) the effectiveness of the proposed DRL framework, 2) the enhanced performance of the DRL optimized trajectory compared to sinusoidal motions, and 3) physical insight on the benefit of DRL optimization.  

\subsection{Whether it can flap: the DRL Training process of different agents } \label{sec:3.1}
\begin{figure}
    \centerline{\includegraphics[width=0.8\textwidth]{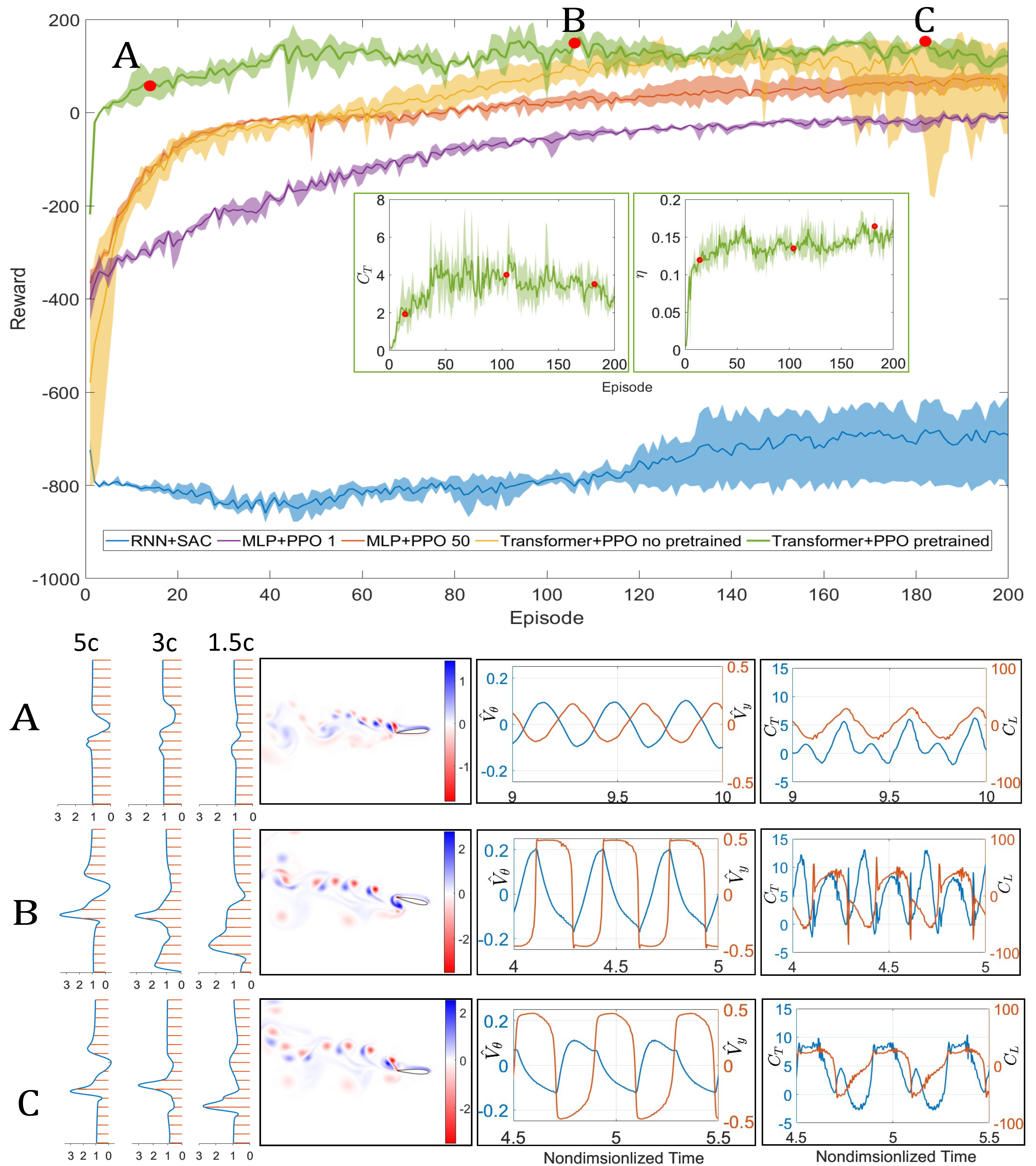}}
    \caption{(Top) Reward over 200 episodes of different combinations of RL algorithms and NN structures. The solid line and shadow represent the mean and variance of three repeated training results. The inset $C_T$ and $\eta$ plots are selected for TPPT agent. Note that the agent of MLP+PPO 1 (50) uses the current observation (50 history data collection) as the state. (Bottom) The mean wake velocity profiles, instantaneous vorticity, the time traces of actions $(\hat{V}_y, \hat{V}_{\theta})$ and forces $(C_T, C_L)$ (from left to right) for three cases denoted as red dots in the top figure: (A) the 14$^{th}$ episode, $\bar{C}_T = 1.93$ and $\eta = 0.12$; (B) the 104$^{th}$ episode, $\bar{C}_T = 4.0$ and $\eta = 0.13$; (C) the 182$^{th}$ episode, $\bar{C}_T = 3.52$ and $\eta = 0.16$.}
    \label{fig:3.1reward}
\end{figure}

To demonstrate the effectiveness and efficiency of the proposed learning framework, we compare in figure \ref{fig:3.1reward} the reward (set $(\beta_C$, $\beta_\mathcal{P}$, $C_T^m$, $C_{\mathcal{P}}^m) = (0.1, 1/3000, 10, 3000)$ in (\ref{eq:r})) of each episode for selected different combinations of RL algorithms and NN structures, including RNN+SAC, Multilayer Perceptron(MLP)+PPO, and Transformer+PPO (TP). 

\begin{figure}
    \centerline{\includegraphics[width=0.8\textwidth]{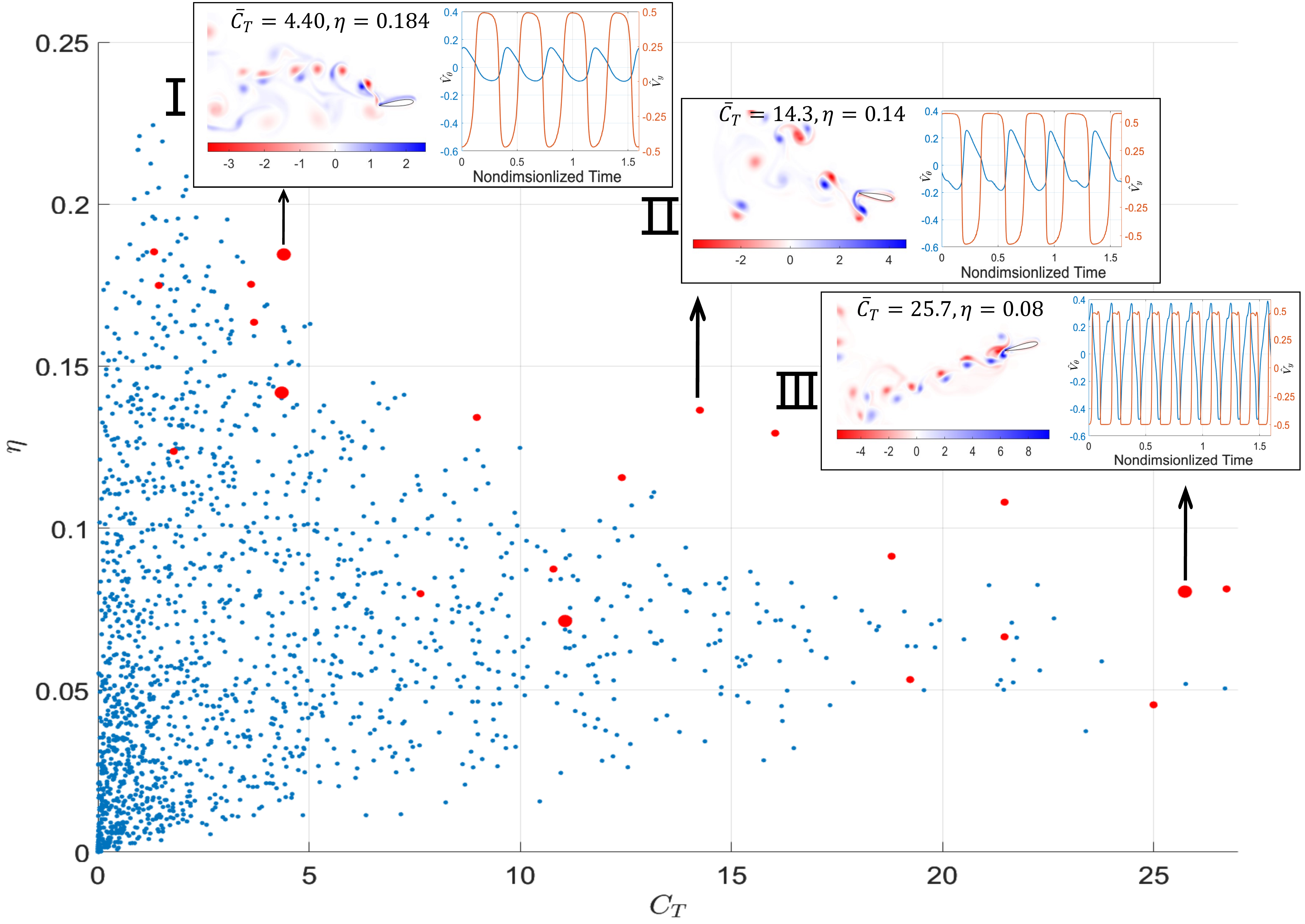}}
    \caption{Scatters of $\bar{C}_T$ and $\eta$ for sinusoidal (blue) and TPPT agent (red) optimized motions. Note that the big red dots have different rewards tuples and the small red dots are acquired by adjusting $\hat{\mathcal{L}}$ after the training is finished. The  inset figures plot instantaneous vorticity, the $\hat{V}_y$ (red) and $\hat{V}_{\theta}$ (blue) time trace of the selected three cases. In scenario \uppercase\expandafter{\romannumeral1}, $\bar{C}_T = 4.40$, $\eta = 0.184$, the reward tuple is $(0.1, \frac{1}{3000}, 10, 3000)$ and $\hat{\mathcal{L}} = [0.5,0.5]$. In scenario \uppercase\expandafter{\romannumeral2}, $\bar{C}_T = 14.3$, $\eta = 0.14$, reward tuple is $(0.1, \frac{1}{3000}, 10, 3000)$ and $\hat{\mathcal{L}} = [0.58,0.58]$. In scenario \uppercase\expandafter{\romannumeral3}, $\bar{C}_T = 25.7$,$\eta = 0.08$, the reward tuple is$(0.1, \frac{1}{3000}, 30, 3000)$ and the $\hat{\mathcal{L}} = [0.5,0.5]$. 
    }
    \label{fig:frontline}
\end{figure}

Figure \ref{fig:3.1reward} (top) shows the two TP agents (green and yellow) manage to learn the nature of oscillatory flapping motion while reaching the highest reward, compared with other selected algorithms. In addition, the TP with pretraining (TPPT) agent reaches convergence faster and with a smaller variance (shadow) among repeated training processes, compared with the agent of TP using random initialization. Furthermore, we inset the $C_T$ and $\eta$ figure for the TPPT agent, showing an increase and convergence of the hydrodynamic performance. 

We select three cases (red dots in figure \ref{fig:3.1reward}) to describe the evolution of the TPPT agent and plot their mean wake velocity profiles, instantaneous vorticity, the time history of actions and forces in figure\ref{fig:3.1reward} (bottom). It first can be seen from case A that the action curve resembles a sinusoidal motion owing to the expert demonstration from the pretraining. Then the action curve in case B becomes non-parametric, though still oscillatory, yet evolves into a period with sudden changes, spikes, and plateaus. In the meantime, comparing the mean velocity profiles in the wake between cases A and B, a stronger jet develops, indicating the hydrodynamic performance of a larger $C_T$ learned by the TPPT agent. Though the reward does not change much as training continues, we observe that action in case C becomes smoother, accompanied by a less spiked force profile. Note that $\eta$ increases by 23$\%$ from case B to C.
 
\subsection{How well it flaps: the hydrodynamic outperformance through reward shaping}

After showing feasibility and efficiency of the TPPT agent, to demonstrate its superior on foil trajectory optimization, we compare in figure \ref{fig:frontline} the hydrodynamic performances of the different TPPT agents optimized trajectories with sinusoidal motion Brute-force (B.F.) $\bar{C}_T, \eta$ search results some of which work as expert policy and initialization of the TPPT agent. 

Figure \ref{fig:frontline} shows the Pareto front formed by B.F. results (small blue dots) where parameters search space are $Sr \in [0.1,0.4]$, $h_0 \in [0.1,0.6]$, $\theta_0 \in [5^{\circ},70^{\circ}]$, $\phi \in [0^{\circ},180^{\circ}]$. In addition, two TPPT agents (big red dots) and four evolved TPPT agents (small red dots) among different training obviously exceed the Pareto front under different reward tuples and $\hat{\mathcal{L}}$ values setting. 

We select three scenarios to clearly show the enhancement of $\bar{C}_T$ and $\eta$ introduced by trajectory change and plot their instantaneous vorticity and the time history of actions. It can be seen that all scenarios' actions are non-sinusoidal, sharing the same features of case C in \ref{sec:3.1}. In scenario \uppercase\expandafter{\romannumeral1}, $\bar{C}_T, \eta$ improves compared to the case C in \ref{sec:3.1}, but action history is different due to total training episode being prolonged. Compared to scenario \uppercase\expandafter{\romannumeral1}, when increase $\hat{\mathcal{L}}$ value, the $\bar{C}_T$ of scenario \uppercase\expandafter{\romannumeral2} get extremely enhanced and amplitude of $\hat{V}_{\theta}$ increase with spike appearance. Compared to scenario \uppercase\expandafter{\romannumeral1}, when increasing the $C_{T}^{m}$ to emphasize $\bar{C}_T$ optimization, the $\bar{C}_T$ of scenario \uppercase\expandafter{\romannumeral3} get further extremely enhanced and it's action frequency and amplitude increase with small oscillation. Note that though $\eta$ decreases from scenario \uppercase\expandafter{\romannumeral1} to scenario \uppercase\expandafter{\romannumeral3}, the overall hydrodynamic performances still outperform the B.F. results.

\subsection{Why it flaps better: the physical insight of DRL optimization strategy}

\begin{figure}
    \centerline{\includegraphics[width=0.9\textwidth]{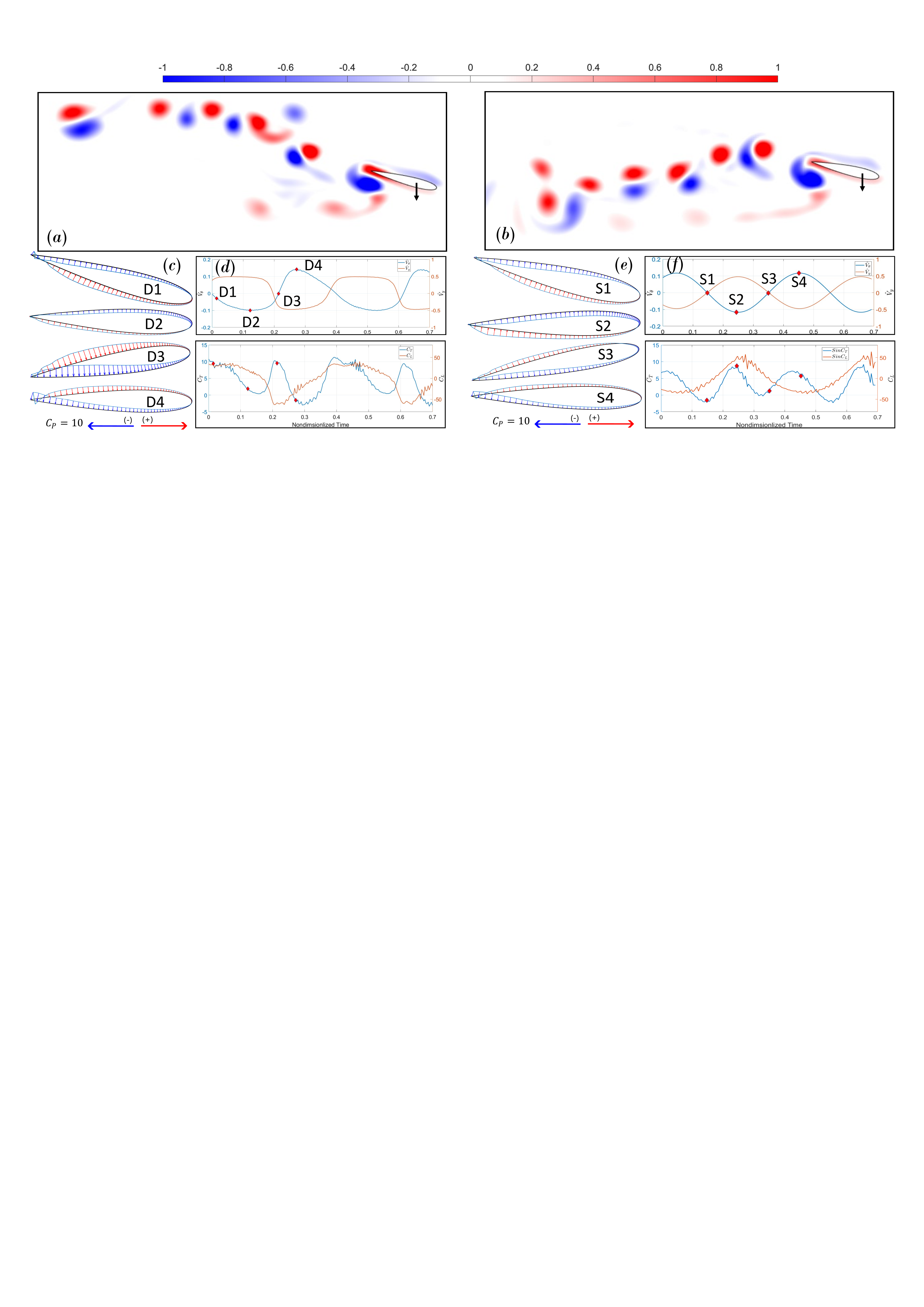}}
    \caption{Instantaneous vorticity, pressure distributions and time trace of action and force over two vortex shedding periods for TPPT scenario \uppercase\expandafter{\romannumeral1} (left)(see Movie 1) and its statistically equivalent sinusoidal motion (right)(see Movie 2). Four foil posture moments are selected to show complete pressure distribution change. For better visualization, the foil shown here is NACA 0020, and the $C_p$ is scaled to 1/10 from its original value.}
    \label{fig:3.2 PA}
\end{figure}

To shed light on the underlying mechanism behind the flapping performance improvement from the TPPT agent, we analyze and plot in figure \ref{fig:3.2 PA} the vorticity ($a$, $b$), pressure distributions around foil($c$, $e$), time history of actions and forces ($d$, $f$) of scenario \uppercase\expandafter{\romannumeral1} (left, from figure \ref{fig:frontline}) and its statistically equivalent (same $S_r$, $h_0$, $\theta_0$ and $\phi$) sinusoidal motion counterpart (right).

In figure4 ($a$, $b$), stronger separated vortices are observed in the whole area. A closer look at the stronger leading-edge vortices reveals that the TPPT-controlled motion generates higher negative pressure alongside forward-facing surfaces, contributing to thrust \citep{batchelor1967introduction}.
The effects of strengthened separated wake are represented quantitatively by pressure distributions around the foil, compared in figure4 ($c$, $e$). 
Based on the orientation of the foil surface, the pressure acting perpendicularly to the surface can contribute to thrust or drag forces \citep{lucas2020airfoil}.
The values of positive and negative pressure around the foil trailing edge are significantly increased in the D3 moment when compared with S3, which functions to increase $C_T$ directly.
Note that the D3 moment is the moment with the maximum acceleration. 
Thus, we hypothesize that under the objectives of maximizing $\bar{C}_T$ and minimizing $\eta$, the TPPT agent senses the separated vortices around the foil body via pressure and takes advantage of them by actively controlling the kinematics and thus improves the hydrodynamic performance. 
This agrees with the hypothesis \citep{muller1997fish} that fish can adjust their kinematics to control near-body flow and improve swimming performance.

\section{Conclusion}
\label{sec:4}

In the present work, we aim to answer whether the DRL agent can learn a reasonable strategy for complex unsteady fluid control problems such as foil flapping, how well it performs compared to the sinusoidal motion, and if so, why the agent can learn better. 

By carefully devising the training framework (TPPT in our case), the agent can outperform the brute-force search of the sinusoidal motion by taking advantage of the vortex-foil interaction and learning coherent non-parametric trajectories. In addition, by adding expert data as initialization, the agent can reach convergence rapidly with the highest reward value.

Furthermore, with a close look into the wake morphology and instantaneous pressure around the foil, the agent can adaptively adjust the statistically similar sinusoidal motion, generating stronger vortices and alternating phases between the motions of the foil and shedding vortices, thus leading to an improvement in hydrodynamic performance.

It is noted that in the current work, we select the simulation environment of low mesh density and the Reynolds number for the proof-of-concept demonstration of DRL with unsteady flapping foil flow control. We believe that our result, for the first time, shows the potential of DRL in complex and time-variant flow control, providing a feasible method to reproduce animal-similar flapping motion and solve other complex flow manipulation tasks.


\section*{Appendix A}

\textbf{POMDP} describes the process of an agent at time $t$ and in state $s_t$ receiving observation $o_t$ with a belief $b$ over the state space, and then taking action $a$ based on policy $\pi(a|o,b)$ with feedback reward $r_t$. Specifically, the POMDP is defined by a tuple $(\mathcal{S}, \mathcal{Z}, \mathcal{A}, O, \mathcal{P}, R, \gamma)$, where $\mathcal{S}, \mathcal{Z}, \mathcal{A}$ are finite sets of state $s$, observation $o$, and action $a$. The transition and observation functions $\mathcal{P}: \mathcal{S} \times \mathcal{A} \rightarrow \Delta(\mathcal{S})$ and $O : \mathcal{S} \times \mathcal{A} \rightarrow \Delta(\mathcal{Z})$ describe the probability of the next state $s_{t+1}$ and observation $o_{t+1}$ in a given state $s_t$ after taking a given action $a_t$. In addition, the reward function $R: \mathcal{S} \times \mathcal{A} \rightarrow \mathbb{R}$ defines the reward received by the agent, and $\gamma \in [0, 1]$ is the discount factor. Therefore, the goal of the agent to find a policy $\pi$ that maximizes the expected discounted sum of rewards over time, subject to the uncertainty of the environment, as follows: 
\begin{equation}
    \max_{\pi} \mathbb{E}_{s_0,a_0, s_1, a_1, \dots} \left[ \sum_{t=0}^{\infty} \gamma^t R(s_t, a_t) \right],
\end{equation}
where in the current problem, $o_t$ is a 24-dimensional array, containing the instantaneous heave and pitch positions and velocities, and pressure measured from 20 sparse sensors around the foil. $a_t$ is a 2-dimensional array of the prescribed pitch and heave velocities determined by the agent.

\textbf{PPO algorithm} follows the actor-critic framework in reinforcement learning. The actor $\pi(a|o,b)$,  parameterized as $\theta$, interacts with the environment, while the critic $V(s)$, parameterized as $\phi$, predicts the onward cumulative reward. For Actor, PPO maximizes a clip objective to penalize changes to the policy that move $r_{t}(\theta)$ far away from the old policy:
\begin{equation}
    L_{actor}(\theta)=\hat{\mathbb{E}}_{t}\left[\min \left(r_{t}(\theta) \hat{A}_{t}, \operatorname{clip}\left(r_{t}(\theta), 1-\epsilon, 1+\epsilon\right) \hat{A}_{t}\right)\right],
\end{equation}
where $r_{t}(\theta)=\frac{\pi_{\theta}\left(a_{t} \mid s_{t}\right)}{\pi_{\theta_{\text {old }}}\left(a_{t} \mid s_{t}\right)}$ denote the probability ratio, $\epsilon$ is a hyper-parameter to constrain the change of policy, and $A_t$ is the advantage to reduce policy gradient variance \citep{sutton2018reinforcement}. For the critic, PPO minimizes the temporal difference loss as:
\begin{equation}
    L_{critic}(\phi) = r_{t}+\gamma V\left(s_{t+1};\phi\right)-V\left(s_{t};\phi\right).
\end{equation}
Therefore, the learning objective for PPO is defined as:
$L(\theta, \phi) = L_{critic}(\phi) + L_{actor}(\theta).$

\textbf{Transformer.} The core of the Transformer is the attention mechanism,  defined as:
\begin{align}
    \text{Attention}(\mathbf{Q},\mathbf{K},\mathbf{V})=\text{softmax}\big(\frac{\mathbf{Q}\mathbf{K}^{T}}{\sqrt{d_{k}}}\big)\mathbf{V},
\end{align}
where $\mathbf{Q}, \mathbf{K}, \mathbf{V}$ are vectors of queries, keys, and values, respectively,  which are learned during training, and $d_k$ is the dimension of $\mathbf{Q}$ and $\mathbf{K}$. In self-attentions,  $\mathbf{Q}, \mathbf{K}, \mathbf{V}$ share the same set of parameters. The attention mechanism allows for the estimation of $P(\mathbf{y}{\mid}\mathbf{x})$ or $P(y_n{\mid}x_1,\dots,x_n)$ without the need for recursive processes, as in RNNs, which results in higher computational efficiency and long-term interaction modelling ability. Our customized Transformer architecture uses the history sequence as the belief about the state, which comprises two primary components: a two-layer encoder and a linear layer as the decoder. Each encoder features a self-attention structure and a feed-forward neural network (FNN) layer. The self-attention structure incorporates two attention heads, a hidden state dimension of 32, and a query dimension of 128. Our experiments are conducted on a server with 2 Nvidia A100 GPU and AMD EPYC 7742 CPU.

\section*{Declaration of Interests}
The authors report no conflict of interest.

\bibliographystyle{jfm}
\bibliography{jfm-instructions}

\end{document}